\documentclass[12pt]{article}
\usepackage{graphics} 
\usepackage{cite}
\usepackage{epsfig}
\newcommand   {\etal}    {{\it et~al.}}
\newcommand  {\APC}      {{\it Adv. Protein Chem.\ }}
\newcommand  {\Bioch}    {{\it Biochemistry\ }}

\newcommand  {\EL}       {{\it Europhys.\ Lett.\ }}
\newcommand  {\JCC}      {{\it J.\ Comput.\ Chem.\ \ }}

\newcommand  {\JCP}      {{\it J.\ Chem.\ Phys.\ }}
\newcommand  {\JMB}      {{\it J.\ Mol.\ Biol.\ }}

\newcommand  {\JSP}      {{\it J.\ Stat.\ Phys.\ }}
 
\newcommand  {\Mac}      {{\it Macromolecules\ }}

\newcommand  {\NSB}      {{\it Nat.\ Struct.\ Biol.\ }}

\newcommand  {\Pro}      {{\it Proteins\ Struct.\ Funct.\ Genet.\ }}
\newcommand  {\ProEng}   {{\it Protein\ Eng.\ }}
\newcommand  {\ProSci}   {{\it Protein\ Sci.\ }}

\newcommand  {\PNAS}     {{\it Proc.\ Natl.\ Acad.\ Sci.\ USA\ }}

\newcommand  {\PRL}      {{\it Phys.\ Rev.\ Lett.\ }}

\newcommand  {\Sci}      {{\it Science\ }}

\newcommand  {\TBS}      {{\it Trends Biochem. Sci.\ }}

\newcommand{\beq}{\begin{equation}}
\newcommand{\eeq}{\end{equation}}
\newcommand{\beqa}{\begin{eqnarray}}
\newcommand{\eeqa}{\end{eqnarray}}
\newcommand{\bea}{\begin{eqnarray}}
\newcommand{\eea}{\end{eqnarray}}
\newcommand   {\ev}[1]   {\langle #1\rangle}
\newcommand   {\Ca}      {C${}_{\alpha}$}
\newcommand   {\Cb}      {C${}_{\beta}$}
\newcommand   {\Cp}      {C${}^{\prime}$}
\newcommand   {\Eloc}    {E_{\mbox{{\scriptsize loc}}}}
\newcommand   {\Cv}      {C_{\mbox{{\scriptsize v}}}}
\newcommand   {\Esa}     {E_{\mbox{{\scriptsize sa}}}}
\newcommand   {\Ehb}     {E_{\mbox{{\scriptsize hb}}}}
\newcommand   {\Eaa}     {E_{\mbox{{\scriptsize AA}}}}
\newcommand   {\ephi}    {\epsilon_\phi}
\newcommand   {\epsi}    {\epsilon_\psi}
\newcommand   {\esa}     {\epsilon_{\mbox{{\scriptsize sa}}}}
\newcommand   {\ehb}     {\epsilon_{\mbox{{\scriptsize hb}}}}
\newcommand   {\eaa}     {\epsilon_{\mbox{{\scriptsize AA}}}}
\newcommand   {\shb}     {\sigma_{\mbox{{\scriptsize hb}}}}
\newcommand   {\saa}     {\sigma_{\mbox{{\scriptsize AA}}}}
\newcommand   {\QFU}     {Q_{\mbox{{\scriptsize FU}}}}
\newcommand   {\QBU}     {Q_{\mbox{{\scriptsize BU}}}}
\newcommand   {\dBU}     {\delta_{\mbox{{\scriptsize BU}}}}
\newcommand   {\dFU}     {\delta_{\mbox{{\scriptsize FU}}}}
\newcommand   {\Rg}      {R_{\mbox{{\scriptsize g}}}}

\input{psfig}

\begin{document}

\begin{flushright}
LU TP 00-22\\
Revised version\\ 
October 29, 2000
\end{flushright}

\vspace{0.4in}

\begin{center}
{\LARGE \bf Three-helix-bundle Protein} 

{\LARGE \bf in a Ramachandran Model}

\vspace{.3in}

\large
Anders Irb\"ack, Fredrik Sjunnesson and
Stefan Wallin\footnote{E-mail: irback,\,fredriks,\,stefan@thep.lu.se}\\   
\vspace{0.10in}
Complex Systems Division, Department of Theoretical Physics\\ 
Lund University,  S\"olvegatan 14A,  S-223 62 Lund, Sweden \\
{\tt http://www.thep.lu.se/tf2/complex/}\\

\vspace{0.3in}	

Submitted to \PNAS

\end{center}
\vspace{0.3in}
\normalsize
Abstract:\\
We study the thermodynamic behavior of a model protein with 
54 amino acids that forms a three-helix bundle in its native state. 
The model contains three types of amino acids and five to six 
atoms per amino acid and has the Ramachandran torsional angles 
$\phi_i$, $\psi_i$ as its degrees of freedom. The force field 
is based on hydrogen bonds and effective hydrophobicity forces. 
For a suitable choice of the relative strength of these interactions, 
we find that the three-helix-bundle protein undergoes an abrupt folding 
transition from an expanded state to the native state. Also shown is 
that the corresponding one- and two-helix segments are less stable
than the three-helix sequence.
 
\newpage

\section{Introduction}
 
It is not yet possible to simulate the formation of proteins' native 
structures on the computer in a controlled way. This goal has been achieved 
in the context of simple lattice and off-lattice models, where typically 
each amino acid is represented by a single interaction site corresponding 
to the \Ca\ atom, and such studies have provided valuable insights into 
the physical principles of protein
folding~\cite{Sali:94,Bryngelson:95,Dill:97,Klimov:98,Nymeyer:98}
and the statistical properties of functional protein 
sequences~\cite{Pande:94,Irback:96}.
However, these models have their obvious limitations. 
Therefore, the search for computationally feasible models 
with a more realistic chain geometry remains a highly relevant task.    
 
In this paper, we discuss a model based on the well-known fact that the
main degrees of freedom of the protein backbone are the Ramachandran
torsional angles $\phi_i,\psi_i$~\cite{Ramachandran:68}. Each amino acid is 
represented by five or six atoms, which makes this model computationally 
slightly more demanding than \Ca\ models. On the other hand, it also 
makes interactions such as hydrogen bonds easier to define. The 
formation of native structure is, in this model, driven by hydrogen-bond 
formation and effective hydrophobicity forces; hydrophobicity is 
widely held as the most important stability factor in 
proteins~\cite{Dill:90,Privalov:92}, and hydrogen bonds are essential 
to properly model the formation of secondary structure.  

In this model, we study in particular a three-helix-bundle protein with 
54 amino acids, which represents a truncated and simplified version of the 
four-helix-bundle protein {\it de novo} designed by Regan and 
DeGrado~\cite{Regan:88}. 
This example was chosen partly because there have been earlier studies 
of similar-sized helical proteins using models at comparable levels of 
resolution~\cite{Rey:93,Guo:96,Zhou:97,Koretke:98,Shea:99,Hardin:99,Takada:99}. 
The behavior of small fast-folding proteins is a current topic in both 
theoretical and experimental research, and a three-helix-bundle protein 
that has been extensively studied both 
experimentally~\cite{Bottomley:94,Bai:97} and 
theoretically~\cite{Zhou:97,Shea:99,Guo:97,Kolinski:98} 
is fragment B of staphylococcal protein A.   

In addition to the three-helix protein, to study size dependence, 
we also look at the behavior of the corresponding one- and two-helix 
segments. By using the method of simulated 
tempering~\cite{Lyubartsev:92,Marinari:92,Irback:95}, a careful 
study of the thermodynamic properties of these different chains
is performed.

Not unexpectedly, it turns out that the behavior of the model
depends strongly on the relative strength of the hydrogen-bond
and hydrophobicity terms. In fact, the situation is somewhat  
reminiscent of what has been found for homopolymers with 
stiffness~\cite{Kolinski:86,Doniach:96,Bastolla:97,Doye:98}, 
with hydrogen bonds playing the role of the stiffness term. 
Throughout this paper, we focus on one specific empirical choice of 
these parameters.

For this choice of parameters, we find 
that the three-helix-bundle protein has the 
following three properties. First, it does form a stable 
three-helix bundle (except for a 2-fold topological degeneracy). 
Second, its folding transition is abrupt, from an expanded 
state to the native three-helix-bundle state. Third, compared to
the one- and two-helix segments, it forms a more stable secondary 
structure. It should be stressed that these properties are found 
without resorting to the popular G\=o approximation~\cite{Go:78}, 
in which interactions that do not favor the desired structure
are ignored.  

\section{The Model}

\begin{figure}
\vspace{0mm}
\hspace{40mm}
\rotatebox{270}{\epsfig{figure=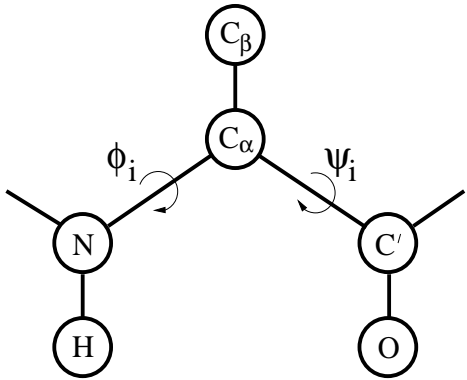,width=5cm,height=6.2cm}}
\vspace{5mm}
\caption{Schematic figure showing the representation of one amino acid.}  
\label{fig:1}
\end{figure}

The model we study is a reduced off-lattice model. The chain   
representation is illustrated in Fig.~\ref{fig:1}. As mentioned in the 
introduction, each amino acid is represented by five or six atoms. 
The three backbone atoms N, \Ca\ and \Cp\ are all included. 
Also included are the H and O atoms shown in Fig.~\ref{fig:1}, which we 
use to define hydrogen bonds. Finally, the side chain is represented by 
a single atom, \Cb, which can be hydrophobic, polar or absent. 
This gives us the following three types of amino acids: A with 
hydrophobic \Cb, B with polar \Cb, and G (glycine) without \Cb.

The H, O and \Cb\ atoms are all attached to the backbone in a rigid 
way. Furthermore, in the backbone, all bond lengths, bond angles and 
peptide torsional angles ($180^\circ$) are held fixed. 
This leaves us with two degrees of freedom per amino acid, the 
Ramachandran torsional angles $\phi_i$ and $\psi_i$ (see Fig.~\ref{fig:1}). 
The parameters held fixed can be found in Table~\ref{tab:1}.

\begin{table}[t]
\begin{center}
\begin{tabular}{llll}
\multicolumn{2}{l}{Bond lengths (\AA)}&
\multicolumn{2}{l}{Bond angles (${}^\circ$)}\\
\hline
N\Ca   & 1.46 & \Cp N\Ca  & 121.7 \\  
\Ca\Cp & 1.52 & N\Ca\Cp   & 111.0 \\
\Cp N  & 1.33 & \Ca\Cp N  & 116.6 \\
NH     & 1.03 & N\Ca\Cb   & 110.0 \\
\Ca\Cb & 1.53 & \Cp\Ca\Cb & 110.0 \\
\Cp O  & 1.23 & & \\
\end{tabular}
\caption{Geometry parameters.}    
\label{tab:1}
\end{center}
\end{table}

Our energy function 
\beq
E=\Eloc+\Esa+\Ehb+\Eaa
\label{e}\eeq
is composed of four terms. 
The local potential $\Eloc$ has a standard form with 3-fold symmetry,
\beq
\Eloc=\frac{\ephi}{2}\sum_i(1 + \cos3\phi_i)
+ \frac{\epsi}{2}\sum_i(1 + \cos3\psi_i)\,.
\eeq
The self-avoidance term $\Esa$ is given by a hard-sphere potential
of the form 
\beq
\Esa=\esa\mathop{{\sum}'}_{i<j}
\bigg(\frac{\sigma_{ij}}{r_{ij}}\bigg)^{12}\,,
\label{sa}\eeq 
where the sum runs over all possible atom pairs except those  
consisting of two hydrophobic \Cb. The hydrogen-bond term $\Ehb$
is given by
\beq
\Ehb= \ehb \sum_{ij}u(r_{ij})v(\alpha_{ij},\beta_{ij})\,,
\label{hb}\eeq
where 
\begin{eqnarray} 
u(r_{ij})&=&  5\bigg(\frac{\shb}{r_{ij}}\bigg)^{12} - 
        6\bigg(\frac{\shb}{r_{ij}}\bigg)^{10}\\
v(\alpha_{ij},\beta_{ij})&=&\left\{ 
        \begin{array}{ll}
 \cos^2\alpha_{ij}\cos^2\beta_{ij} & \ \alpha_{ij},\beta_{ij}>90^{\circ}\\
 0                      & \ \mbox{otherwise}
         \end{array} \right. 
\end{eqnarray}
In Eq.~\ref{hb} $i$ and $j$ represent H and O atoms, respectively, 
and $r_{ij}$ denotes the HO distance, $\alpha_{ij}$ the NHO 
angle, and $\beta_{ij}$ the HO\Cp\ angle. Any HO pair can form a 
hydrogen bond. The last term in Eq.~\ref{e}, the hydrophobicity 
term $\Eaa$, has the form
\beq
\Eaa=\eaa\sum_{i<j}\bigg[
\bigg(\frac{\saa}{r_{ij}}\bigg)^{12}
-2\bigg(\frac{\saa}{r_{ij}}\bigg)^6\,\bigg]\,,
\eeq
where both $i$ and $j$ represent hydrophobic \Cb. To speed up the simulations, 
a cutoff radius $r_c$ is used,\footnote{The cutoff procedure is
$f(r)\mapsto\tilde f(r)$ where $\tilde f(r)=f(r)-f(r_c)-(r-r_c)f^\prime(r_c)$
if $r<r_c$ and $\tilde f(r)=0$ otherwise.} which is 4.5\AA\ for $\Esa$ 
and $\Ehb$, and 8\AA\ for $\Eaa$.

In this energy function, roughly speaking, the first two terms,
$\Eloc$ and $\Esa$, enforce steric constraints, 
whereas the last two terms, $\Ehb$ and $\Eaa$, are the 
ones responsible for stability. 
Force fields similar in spirit, emphasizing hydrogen bonding 
and hydrophobicity, have been used with some success to predict 
structures of peptides~\cite{Ishikawa:99} and small helical 
proteins~\cite{Koretke:98}. 

\begin{figure}
\vspace{-40mm}
\mbox{
  \hspace{-30mm}
  \psfig{figure=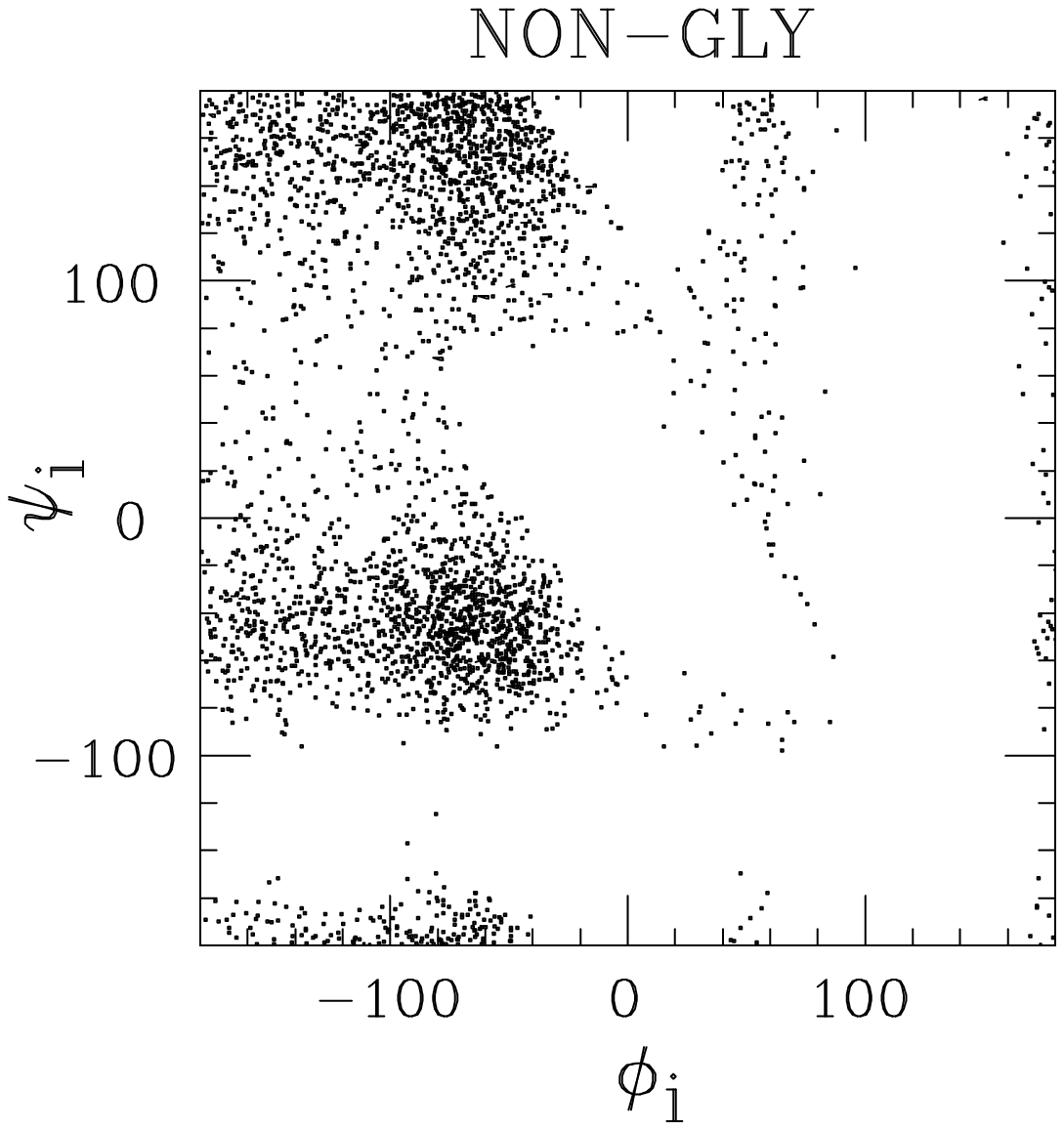,width=10cm,height=14.1cm}
  \hspace{-30mm}
  \psfig{figure=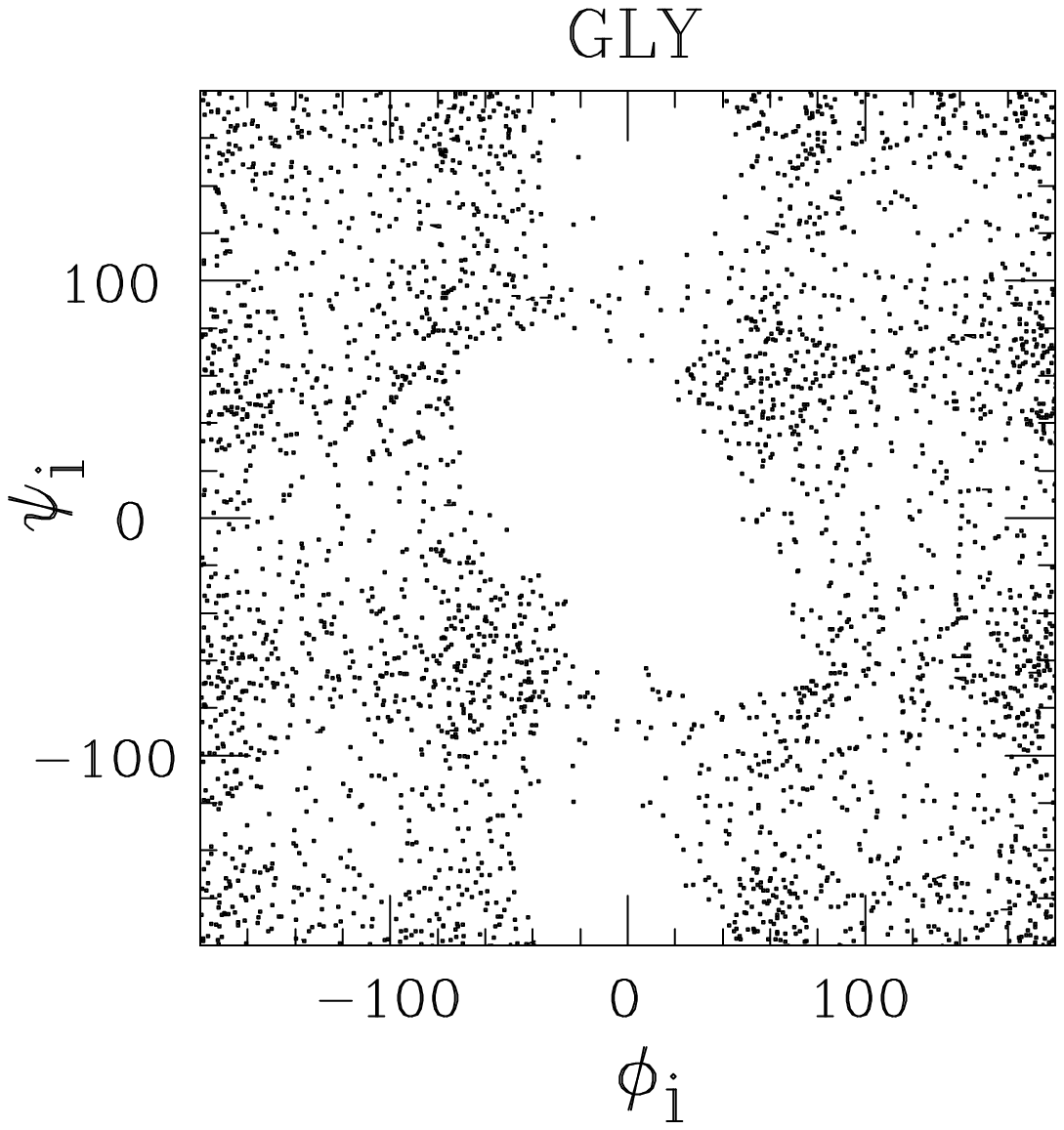,width=10cm,height=14.1cm}
}
\vspace{-40mm}
\caption{$\phi_i,\psi_i$ scatter plots for nonglycine and glycine, as obtained
by simulations of the chains GXG for X=A/B and X=G, respectively, at $kT=0.625$
(shown is $\phi_i,\psi_i$ for X).} 
\label{fig:2}
\end{figure}

The parameters of our energy function were determined largely by
trial and error. The final parameters are listed in 
Table~\ref{tab:2}. The parameters $\sigma_{ij}$ of Eq.~\ref{sa} 
are given by 
\[
\sigma_{ij}=\sigma_{i}+\sigma_j+\Delta\sigma_{ij}\,,
\] 
where $\sigma_i,\sigma_j$ can be found in Table~\ref{tab:2}, and 
$\Delta\sigma_{ij}$ is zero except for \Cb\Cp, \Cb N and \Cb O pairs 
that are connected by three covalent bonds. In these three cases, we put 
$\Delta\sigma_{ij}=0.625$\AA . This could equivalently be described     
as a change of the local $\phi_i$ and $\psi_i$ potentials. In 
Fig.~\ref{fig:2}, we show $\phi_i,\psi_i$ scatter plots for nonglycine (A 
and B) and glycine for our final parameters, which are in
good qualitative agreement with the $\phi_i,\psi_i$ distributions
of real proteins~\cite{Ramachandran:68,Zimmerman:77}.   

Finally, we determined the strengths of the hydrogen-bond and 
hydrophobicity terms on the basis of the resulting overall thermodynamic
behavior of the three-helix sequence.  For this purpose, we performed a 
set of trial runs for fixed values of the other parameters. An alternative 
would have been to use the method of Shea~\etal~\cite{Shea:98}. 
The result of our empirical determination of $\ehb$ and $\eaa$ does 
not seem unreasonable; at the folding temperature of the three-helix 
sequence (see below), we get $\ehb/kT\approx4.3$ and $\eaa/kT\approx3.4$.

\begin{table}[t]
\begin{center}
\begin{tabular}{ccccccccccccc}
\multicolumn{5}{c}{}&\multicolumn{6}{c}{$\sigma_i$(\AA)}\\
\cline{6-11}
$\ephi$ & $\epsi$ & $\esa$ & $\ehb$ & $\eaa$ & N & \Ca & \Cp & H & \Cb & O &
$\shb$ (\AA) & $\saa$ (\AA) \\
\hline  
1 & 1 & 0.0034 & 2.8 & 2.2 & 1.65 & 1.85 & 1.85 & 1.0 & 2.5 & 1.65 & 2.0 & 5.0
\end{tabular}
\caption{Parameters of the energy function. Energies are 
in dimensionless units, in which the folding transition occurs 
at $kT\approx0.65$ for the three-helix-bundle protein (see below).} 
\label{tab:2}
\end{center}
\end{table}

In this model, we study the three sequences shown in Table~\ref{tab:3},  
which contain 16, 35 and 54 amino acids, respectively. Following the 
strategy of Regan and DeGrado~\cite{Regan:88}, the A and B amino acids 
are distributed along the sequence 1H in such a way that this segment 
can form a helix with all hydrophobic amino acids on the same side. 
The sequence 3H, consisting of three such stretches of As and Bs 
plus two GGG segments, is meant to form a three-helix bundle. 
This particular sequence was recently studied by 
Takada~\etal~\cite{Takada:99}, who used a more elaborate model
with nonadditive forces.  

\begin{table}
\begin{center}
\begin{tabular}{ll}
1H:& BBABBAABBABBAABB\\
2H:& 1H--GGG--1H\\
3H:& 1H--GGG--1H--GGG--1H
\end{tabular}
\caption{The sequences studied.} 
\label{tab:3}
\end{center}
\end{table}
 
\section{Results}

To study the thermodynamic behavior of the chains described in the
previous section, we use the method of simulated tempering.
This means that we first select a set of allowed temperatures and 
then perform simulations in which the temperature is a dynamical 
variable. This is done to speed up low-temperature simulations. In addition, 
it provides a convenient method for calculating free energies.  

\begin{figure}
\vspace{-33mm}
\hspace{20mm}
\psfig{figure=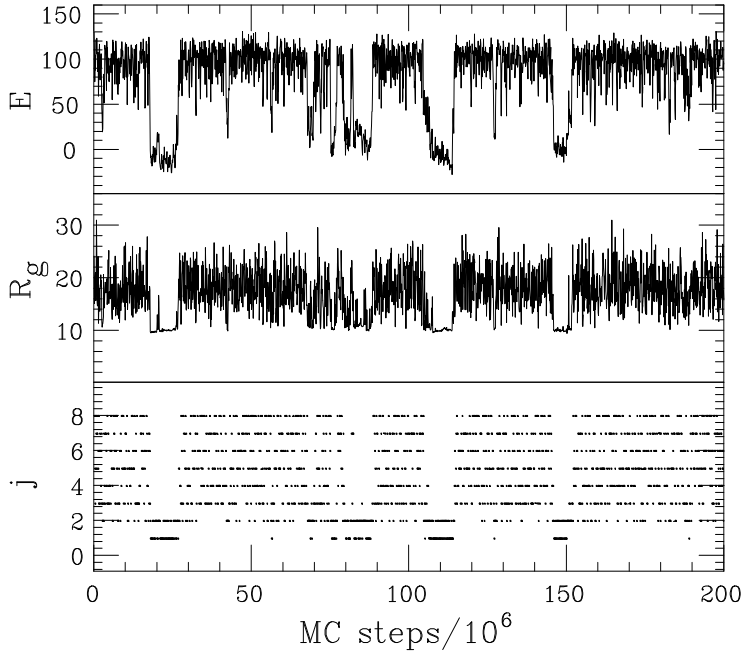,width=10cm,height=14.1cm}
\vspace{-17mm}
\caption{Monte Carlo evolution of the energy and radius of gyration  
in a typical simulation of the three-helix sequence. The bottom panel shows 
how the system jumps between the allowed temperatures $T_j$, which
are given by 
$T_j=T_{\min}(T_{\max}/T_{\min})^{(j-1)/(J-1)}$~\protect\cite{Hansmann:97} 
with $kT_{\min}=0.625$, $kT_{\max}=0.9$ and $J=8$. The temperature $T_{\min}$
is chosen to lie just below the collapse transition, whereas $T_{\max}$ is well 
into the coil phase (see Fig.~\protect\ref{fig:4}).}   
\label{fig:3}
\end{figure}

An example of a simulated-tempering run is given in Fig.~\ref{fig:3},   
which shows the Monte Carlo evolution of the energy $E$ and 
radius of gyration $\Rg$ (calculated over all backbone atoms) in a 
simulation of the three-helix sequence.
Also shown, bottom panel, is how the system jumps between the 
different temperatures. Two distinct types of behavior can be 
seen. In one case, $E$ is high, fluctuations in size are large, 
and the temperatures visited are high. In the other case, $E$ is low, 
the size is small and almost frozen, and the temperatures visited
are low. Interesting to note is that there is one temperature, the
next-lowest one, which is visited in both cases. Apparently, both
types of behavior are possible at this temperature.   

\begin{figure}
\vspace{-42mm}
\mbox{
  \hspace{-51mm}
  \psfig{figure=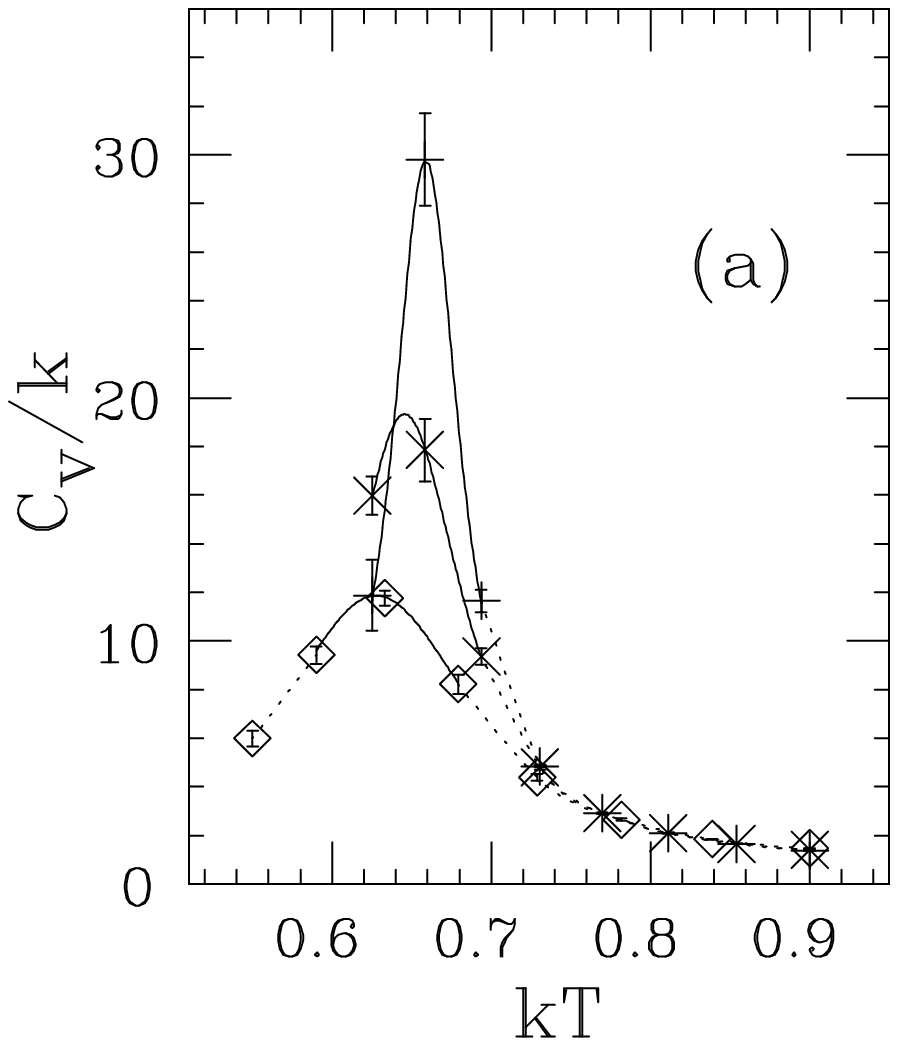,width=10cm,height=14.1cm}
  \hspace{-51mm}
  \psfig{figure=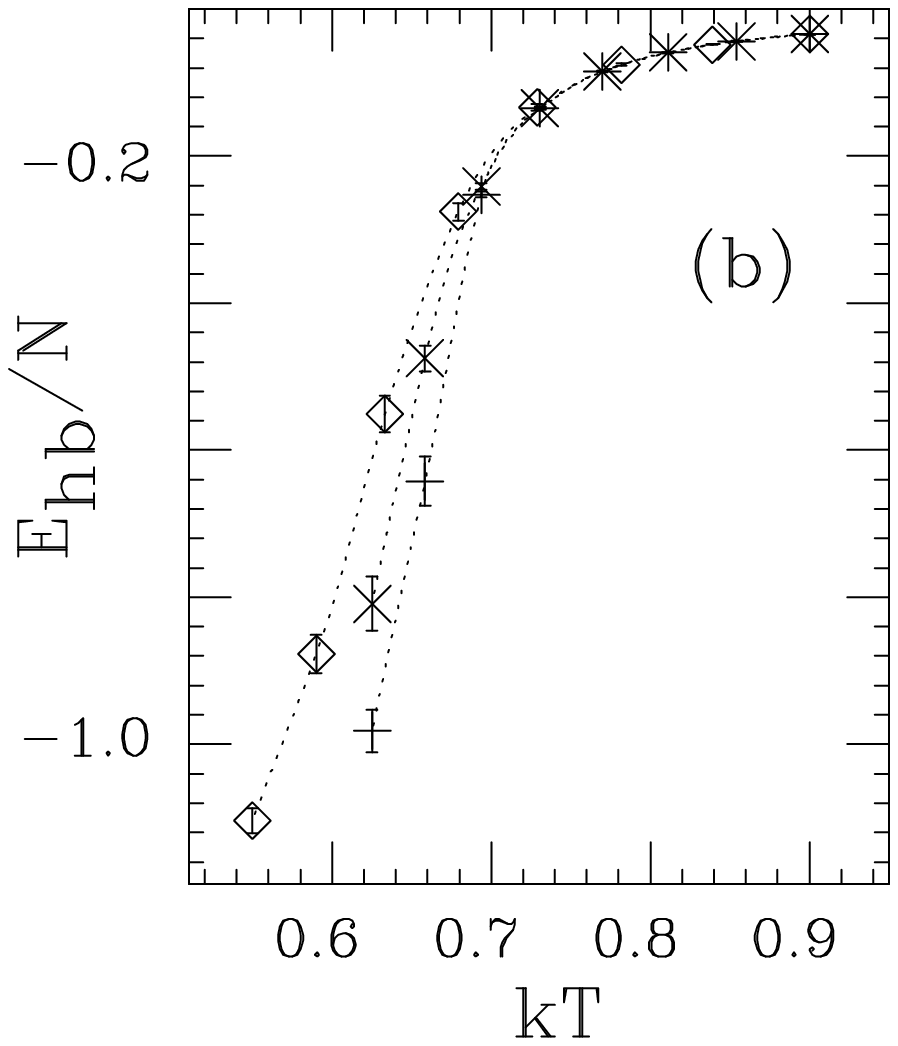,width=10cm,height=14.1cm}
  \hspace{-51mm}
  \psfig{figure=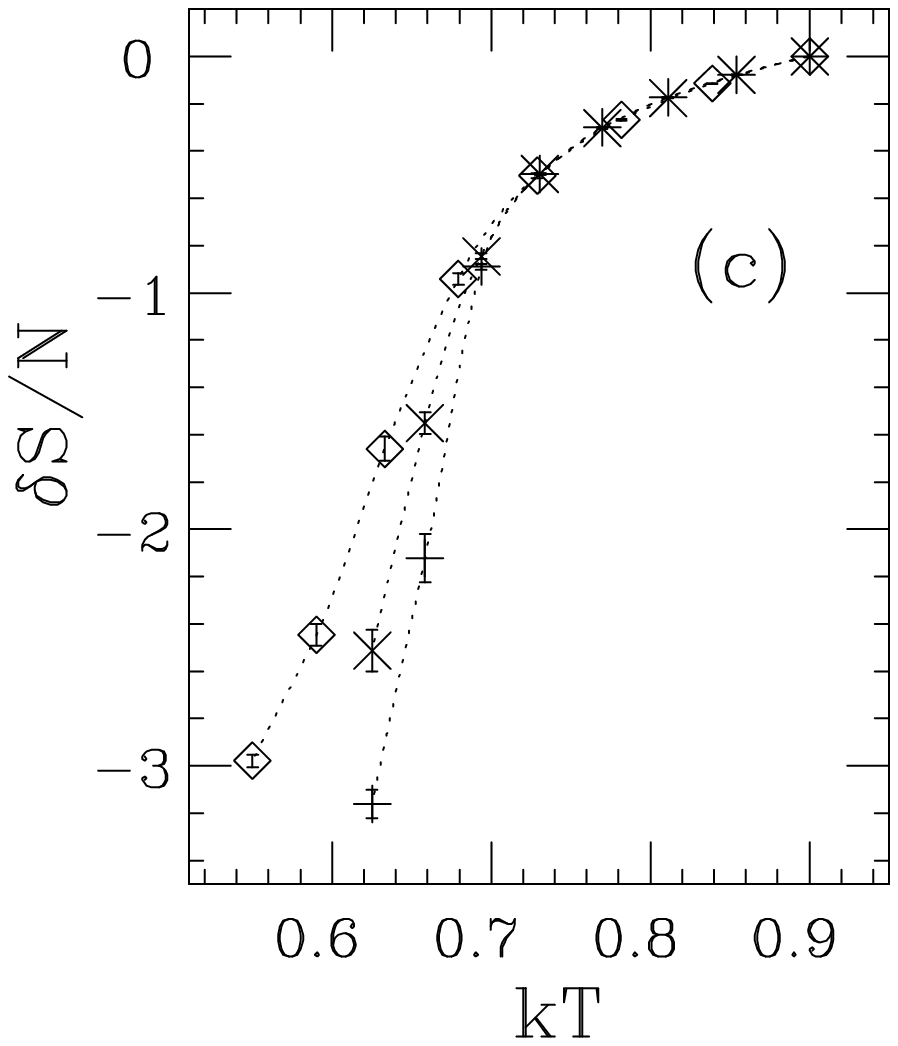,width=10cm,height=14.1cm}
}
\vspace{-47mm}
\caption{Thermodynamic functions against temperature for the 
sequences 1H ($\diamond$), 2H ($\times$) and 3H ($+$) in  
Table~\ref{tab:3}. (a) Specific heat $\Cv=(\ev{E^2}-\ev{E}^2)/NkT^2$,
$N$ being the number of amino acids. 
(b) Hydrogen-bond energy per amino acid, $\Ehb/N$.
(c) Chain entropy per amino acid, $\delta S/N=[S-S(kT=0.9)]/N$.    
The full lines in (a) represent single-histogram 
extrapolations~\protect\cite{Ferrenberg:88}. Dotted lines are 
drawn to guide the eye.}
\label{fig:4}
\end{figure}

In Fig.~\ref{fig:4}a we show the specific heat as a function of 
temperature for the one-, two- and three-helix sequences. 
A pronounced peak can be seen that gets stronger with 
increasing chain length. In fact, the increase in 
height is not inconsistent with a linear dependence on chain
length, which is what one would have expected if it had been a 
conventional first-order phase transition with a latent heat.
       
Our results for the radius of gyration (not displayed) show that 
the specific heat maximum can be viewed as the collapse temperature. 
The specific heat maximum is also where hydrogen-bond formation 
occurs, as can be seen from Fig.~\ref{fig:4}b. Important to note 
in this figure is that the decrease in hydrogen-bond energy 
{\it per amino acid} with decreasing temperature is most rapid for 
the three-helix sequence, which implies that, compared to the shorter ones, 
this sequence forms more stable secondary structure. The results 
for the chain entropy shown in Fig.~\ref{fig:4}c provide further 
support for this; the entropy loss per amino acid with decreasing 
temperature is largest for the three-helix sequence. 

It should be stressed that the character of the collapse transition
depends strongly on the relative strength of the hydrogen-bond
and hydrophobicity terms. Figure~\ref{fig:4} shows that the transition is
very abrupt or ``first-order-like'' for our choice $(\ehb,\eaa)=(2.8,2.2)$.
A fairly small decrease of $\ehb/\eaa$ is sufficient to get a very 
different behavior with, for example, a much weaker peak in the specific 
heat. In this case, the chain collapses to a molten globule without specific 
structure rather than to a three-helix bundle. A substantially weakened 
transition was observed for $\ehb=\eaa=2.5$. 
If, on the other hand, $\ehb/\eaa$ is too large, 
then it is evident that the chain will form one long helix instead
of a helical bundle.

\begin{figure}
\vspace{-00mm}
\mbox{
 \hspace{40mm}
 \epsfig{figure=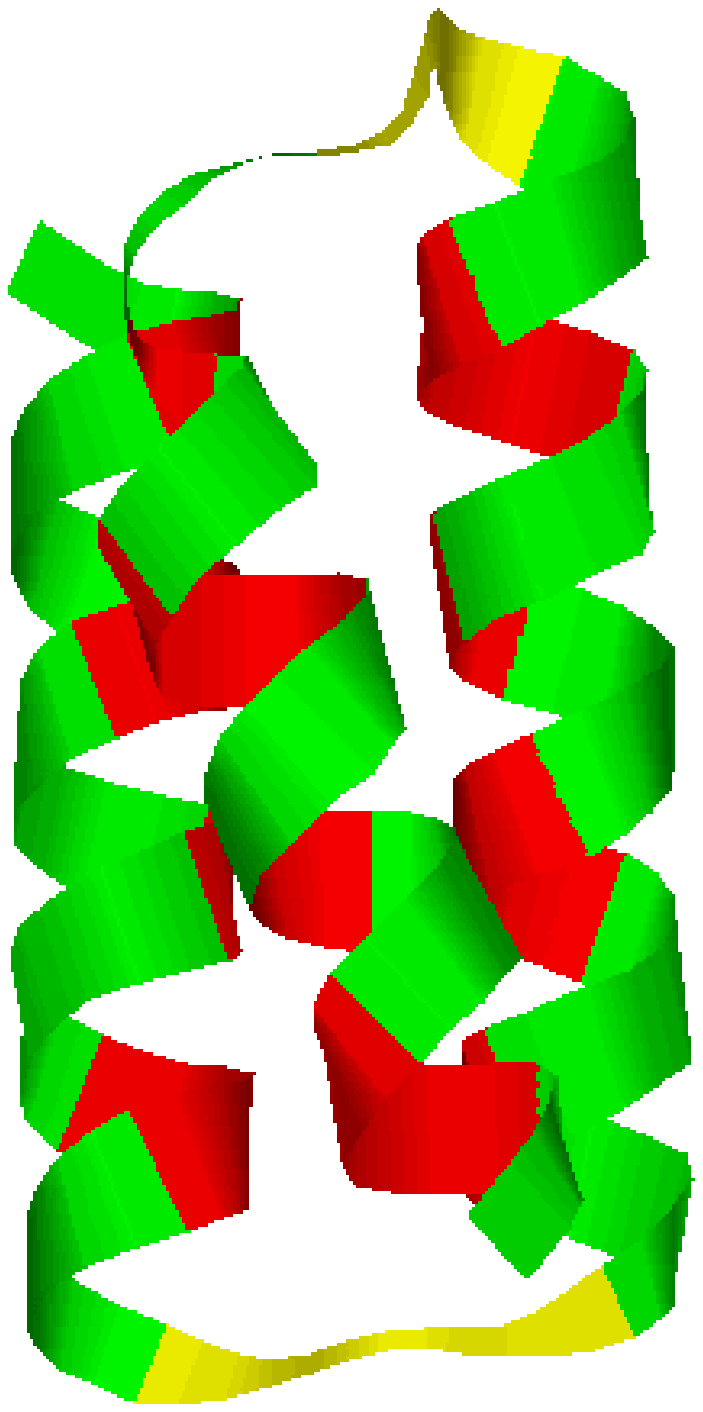,width=1.53cm,height=2.97cm}
 \hspace{30mm}
 \epsfig{figure=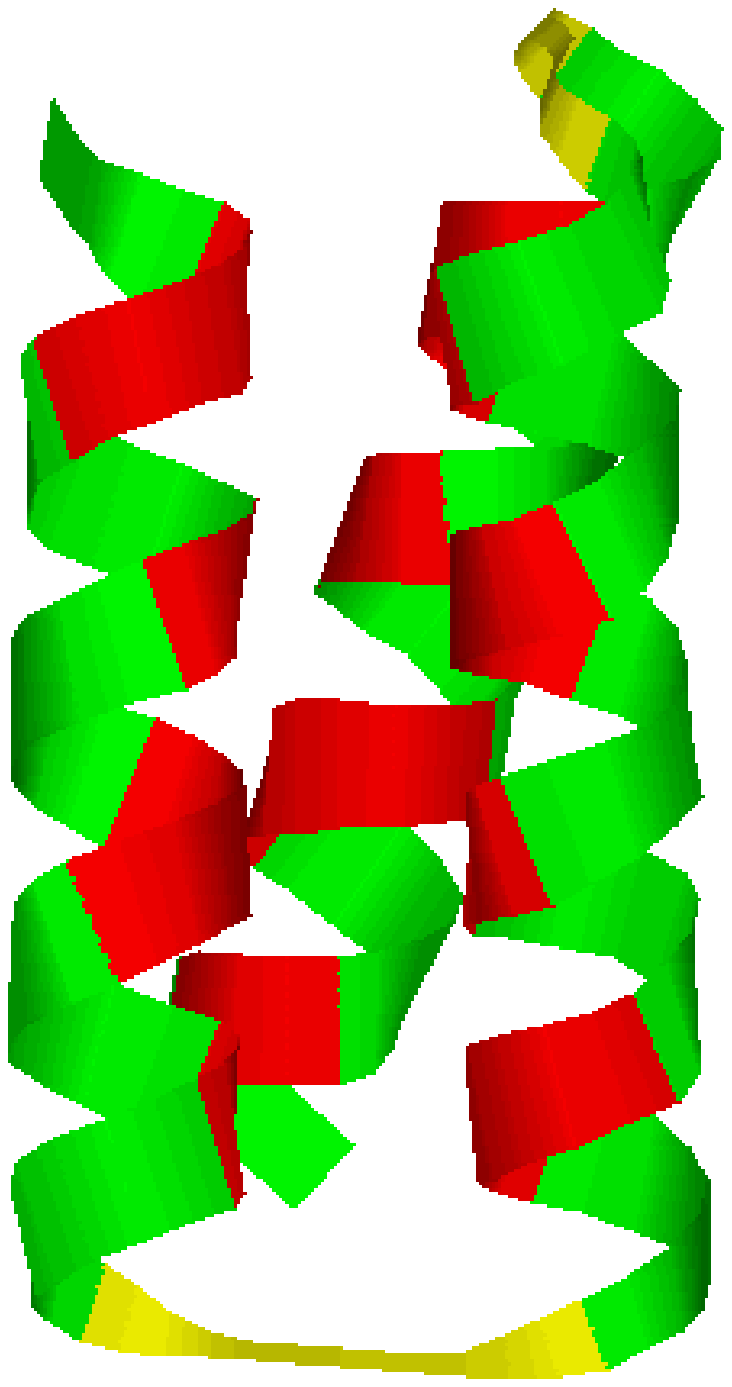,width=1.53cm,height=2.97cm}
}
\caption{Representative low-temperature structures, FU and BU, respectively.
Drawn with RasMol~\protect\cite{Sayle:95}.}  
\label{fig:5}
\end{figure}

We now turn to the three-dimensional structure of the three-helix 
sequence in the collapsed phase. It turns out that it does form a 
three-helix bundle. This bundle can have two distinct topologies: if we 
let the first two helices form a U, then the third helix can be either 
in front of or behind that U.  The model is, not unexpectedly, 
unable to discriminate between these two possibilities. To characterize 
low-temperature conformations, we therefore determined two representative 
structures, one for each topology, which, following~\cite{Takada:99}, 
are referred to as FU and BU, respectively. These structures
are shown in Fig.~\ref{fig:5}. They were generated by 
quenching a large number of low-$T$ structures to 
zero temperature, and we feel convinced that they provide good 
approximations of the energy minima for the respective topologies.
Given an arbitrary conformation, we then measure the root-mean-square 
distances $\delta_i$ ($i=$\,FU,\,BU) to these two structures (calculated
over all backbone atoms). These distances are 
converted into similarity parameters $Q_i$ by using  
\beq
Q_i=\exp(-\delta_i^2/100\mbox{\AA}^2)\,.
\label{sim}\eeq
             
\begin{figure}
\vspace{-40mm}
\mbox{
  \hspace{-30mm}
  \psfig{figure=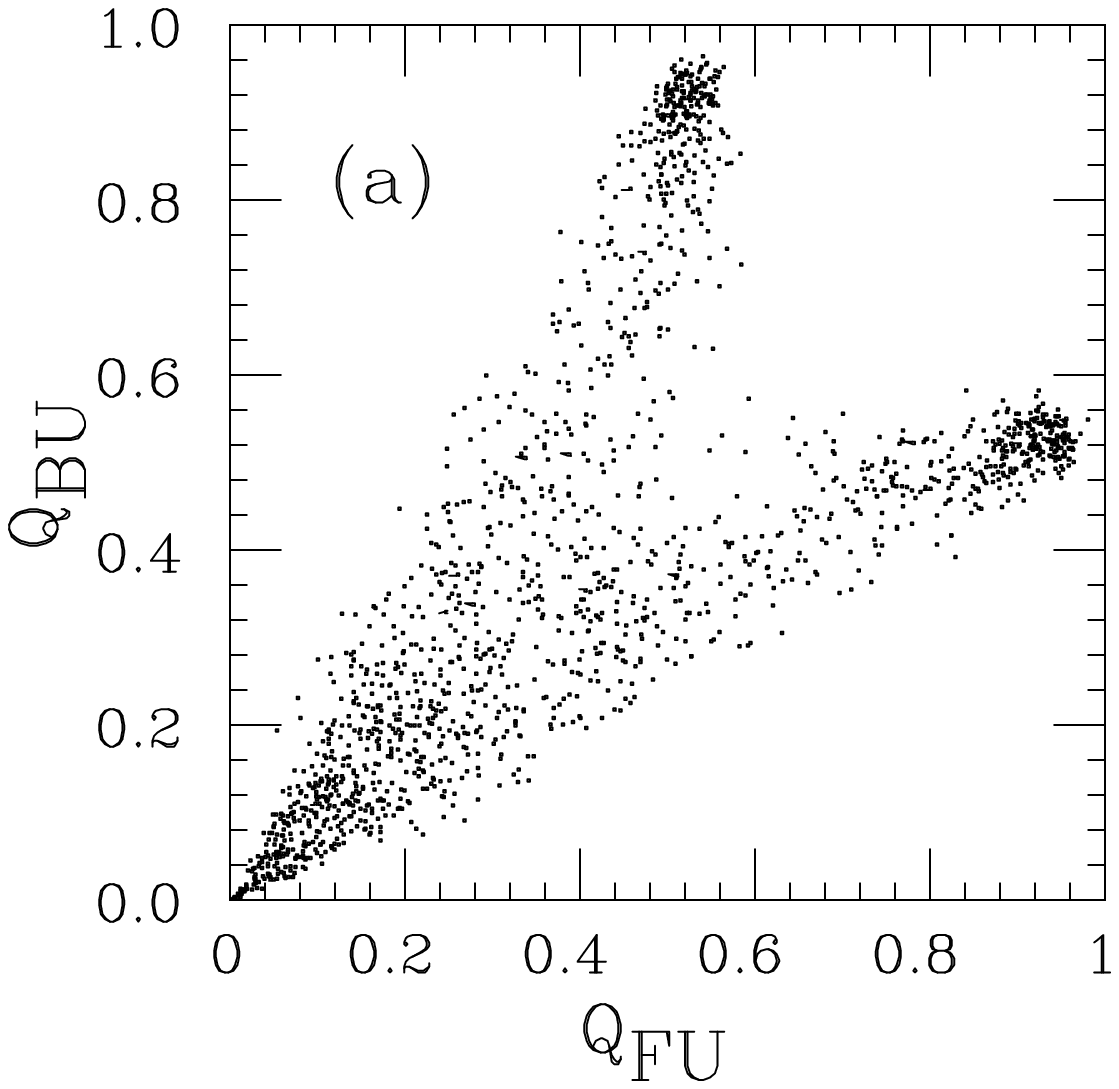,width=10cm,height=14.1cm}
  \hspace{-30mm}
  \psfig{figure=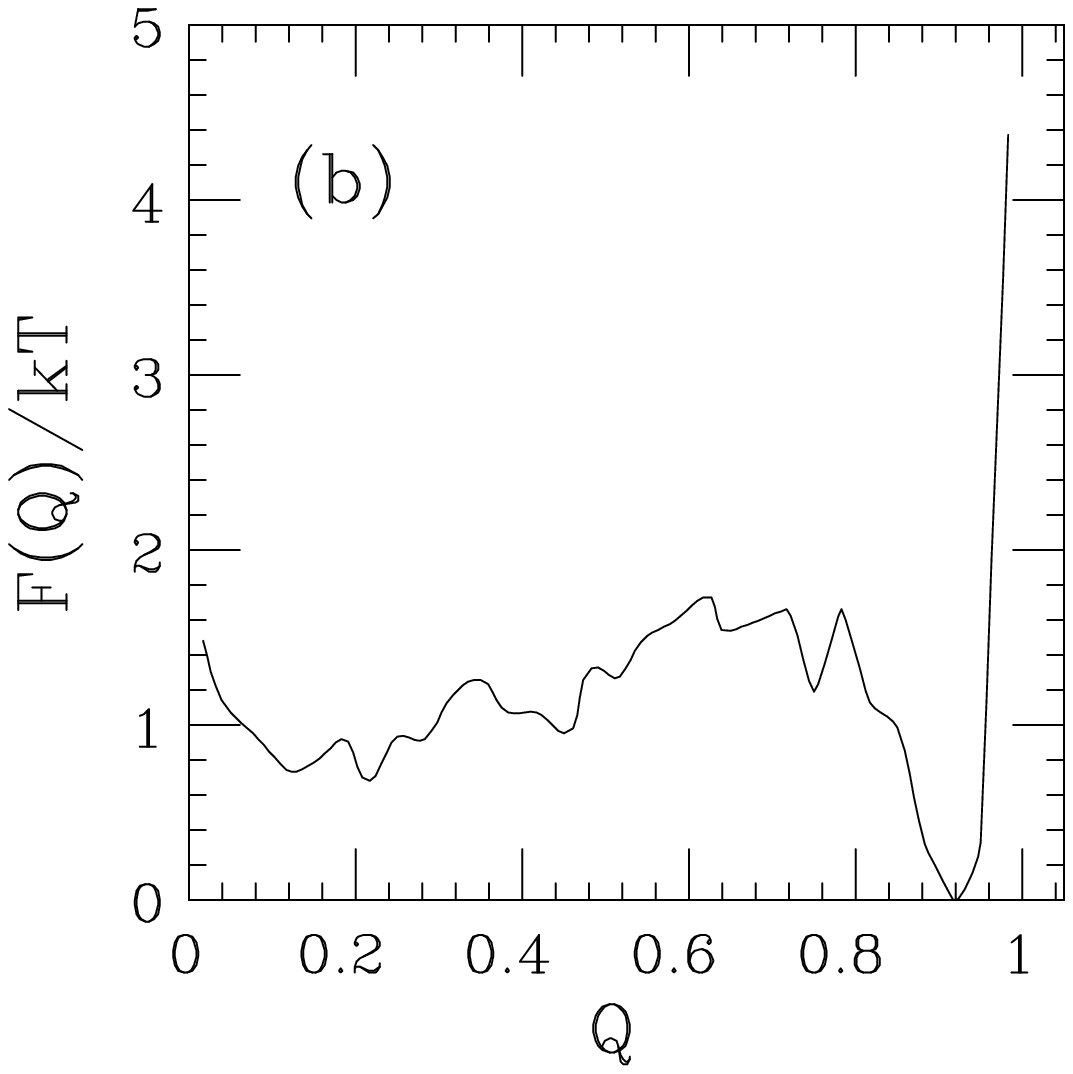,width=10cm,height=14.1cm}
}
\vspace{-40mm}
\caption{(a) $\QFU,\QBU$ (see Eq.~\protect\ref{sim}) scatter plot 
at the specific heat maximum ($kT=0.658$). 
(b) Free energy $F(Q)$ as a function of $Q=\max(\QFU,\QBU)$ at 
the same temperature.}     
\label{fig:6}
\end{figure}

At temperatures above the specific heat maximum, both $Q_i$ tend 
to be small. At temperatures below this point, the system is 
found to spend most of its time close to one or the other of the  
representative structures; either $\QFU$ or $\QBU$ is close to 1. 
Finally, at the peak, all three of these regions in the $\QFU,\QBU$
plane are populated, as can be seen from Fig.~\ref{fig:6}a. In particular,
this implies that the folding transition coincides with the 
specific heat maximum. 

The folding transition can be described in terms of a single 
``order parameter'' by taking $Q=\max(\QFU,\QBU)$ as a  
measure of nativeness. Correspondingly, we put 
$\delta=\min(\dFU,\dBU)$. In Fig.~\ref{fig:6}b, 
we show the free-energy profile $F(Q)$ at the folding temperature. 
The free energy has a relatively sharp minimum at $Q\approx0.9$, 
corresponding to $\delta\approx3$\AA. This is followed by a weak 
barrier around $Q=0.7$, corresponding to $\delta\approx6$\AA. 
Finally, there is a broad minimum at small $Q$, where $Q=0.2$ corresponds 
to $\delta\approx13$\AA.   

\begin{figure}
\vspace{-40mm}
\mbox{
  \hspace{-30mm}
  \psfig{figure=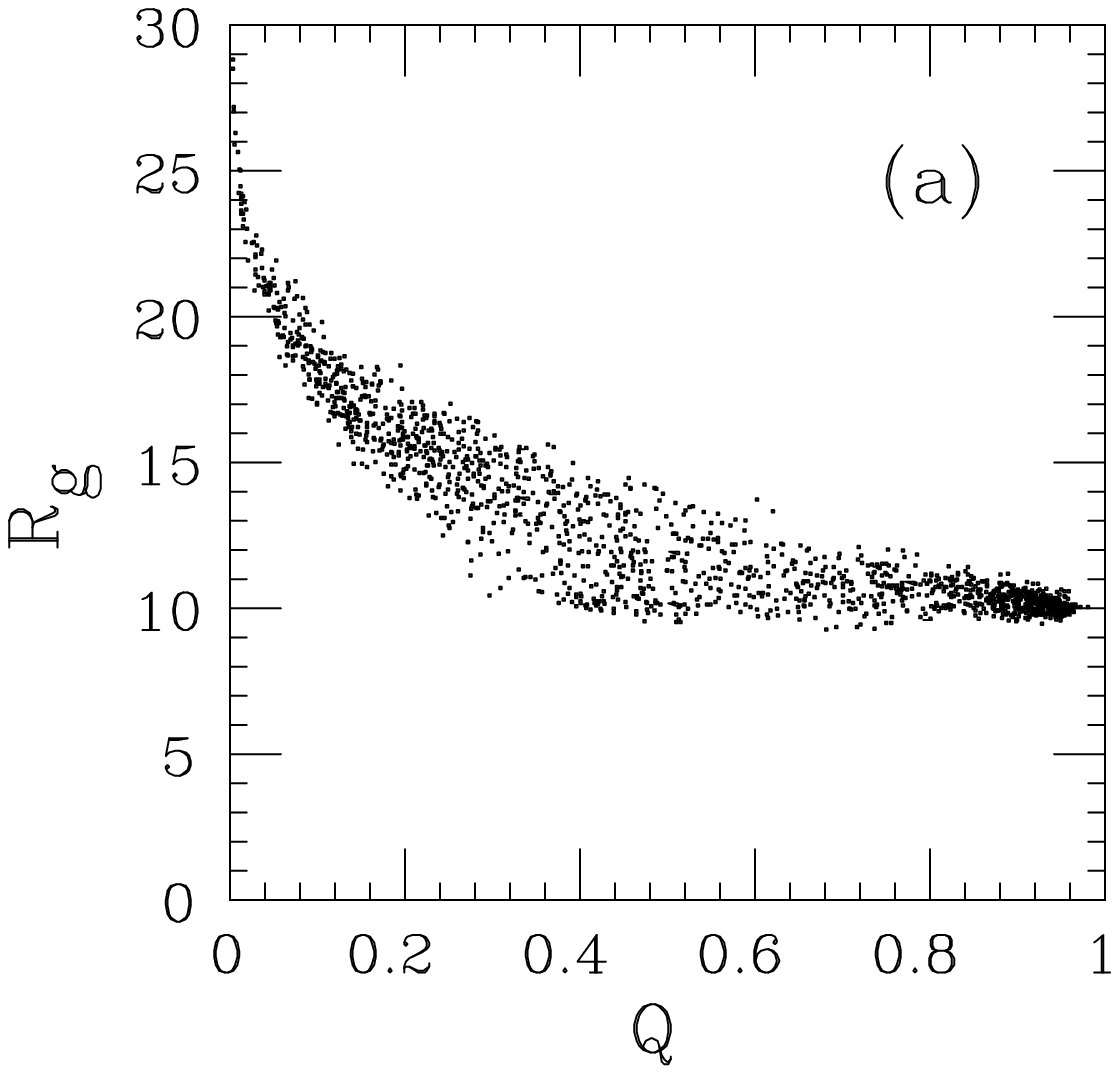,width=10cm,height=14.1cm}
  \hspace{-30mm}
  \psfig{figure=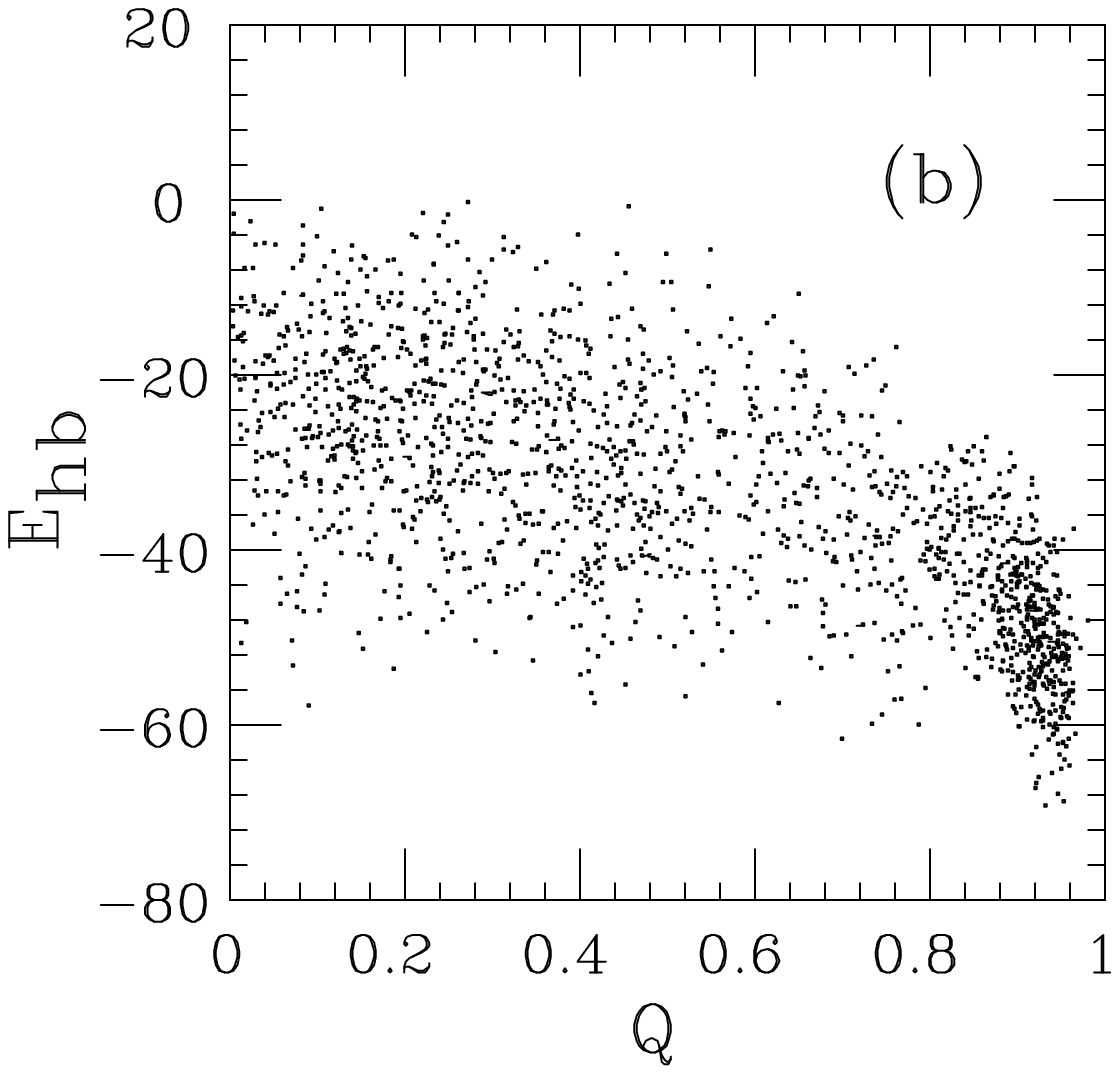,width=10cm,height=14.1cm}
}
\vspace{-40mm}
\caption{(a) $Q,\Rg$ and (b) $Q,\Ehb$ scatter plots at the 
folding temperature ($kT=0.658$).} 
\label{fig:7}
\end{figure}

What does the nonnative population at the folding temperature 
correspond to in terms of $\Rg$ and $\Ehb$? This can be seen from 
the $Q,\Rg$ and $Q,\Ehb$ scatter plots in Fig.~\ref{fig:7}.
These plots show that the low-$Q$ minimum of $F(Q)$ corresponds to 
expanded structures with a varying but not high 
secondary-structure content. Although a detailed kinetic study 
is beyond the scope of this paper, we furthermore note that the 
free-energy surfaces corresponding to the distributions in 
Fig.~\ref{fig:7} are relatively smooth. Consistent with that, we 
found that standard fixed-temperature Monte Carlo simulations were able to 
reach the native state, starting from random coils. 

Let us finally mention that we also performed simulations of some 
random sequences with the same length and composition as the 
three-helix sequence. The random sequences did not form stable 
structures and collapsed more slowly with decreasing temperature
than the designed three-helix sequence.  

\section{Summary and Outlook}
 
We have studied a reduced protein model where the formation 
of native structure is driven by a competition between hydrogen 
bonds and effective hydrophobicity forces. Using this force field, 
we find that the three-helix-bundle protein studied 
has the following properties: 
\begin{itemize}
\item It does form a stable three-helix-bundle state, except 
for a 2-fold topological degeneracy.
\item It undergoes an abrupt folding transition from an expanded
state to the native state. 
\item It forms more stable secondary structure than the
corresponding one- and two-helix segments.
\end{itemize}
An obvious question that remains to be addressed is what is 
needed to lift the topological degeneracy. Not obvious, however, 
is whether this question should be addressed at the present 
level of modeling, before including full side chains.     
     
A first-order-like 
folding transition that takes the system directly from the 
unfolded state to the native one is what one expects
for small fast-folding proteins. For the model to
show this behavior, careful tuning of the relative
strength of the hydrogen-bond and hydrophobicity terms, $\ehb/\eaa$,
is required. This $\ehb/\eaa$ dependence may at first glance seem unwanted 
but is not physically unreasonable; $\ehb$ can be thought of partly as a 
stiffness parameter, and chain stiffness has important implications  
for the phase structure, as shown by recent work on 
homopolymers~\cite{Kolinski:86,Doniach:96,Bastolla:97,Doye:98}.
Note also that incorporation of full side chains makes the chains 
intrinsically stiffer, which might lead to a weaker    
$\ehb/\eaa$ dependence. 

Our three-helix sequence has previously been studied by  
Takada~\etal~\cite{Takada:99}, who used a more elaborate force
field. It was suggested that it is essential to use 
context-dependent hydrogen bonds for the three-helix-bundle 
protein to make more stable secondary structure than its one-helix fragments. 
Our model shows this behavior, although its hydrogen bonds are 
context-independent.

Let us finally stress that we find a first-order-like
folding transition without using the G\=o approximation. 
Evidence for first-order-like folding transitions has been found 
for proteins with similar lengths in some \Ca\ 
models~\cite{Zhou:97,Nymeyer:98,Shea:99,Shea:98}, but these 
studies use this approximation. 

\section*{Acknowledgements}

This work was in part supported by the Swedish Foundation for Strategic 
Research.

\newpage

%\bibliography{/home/people/irback/tex/bib/refs}
%\bibliographystyle{unsrt}

\end{document}